

\magnification=1200
\hoffset=-.1in
\voffset=-.2in
\font\cs=cmcsc10
\font\eightrm=cmr8
\vsize=7.5in
\hsize=5.6in
\tolerance 10000
\def\ll{\left\langle}
\def\rr{\right\rangle}
\def\Tr{\,{\rm Tr}\,}
\def\svec#1{\skew{-2}\vec#1}

\def\footstrut{\baselineskip 12pt}
\baselineskip 12pt plus 1pt minus 1pt
\pageno=0
\centerline{\bf LATTICE CALCULATION OF QCD}
\medskip
\centerline{{\bf VACUUM CORRELATION FUNCTIONS }\footnote{*}{This
work is supported in part by funds
provided by the U. S. Department of Energy (D.O.E.) under contracts
\#DE-AC02-76ER03069 and \#DE-FG06-88ER40427, and the National Science
Foundation under grant \#PHY~88-17296.}}
\vskip 24pt
\centerline{M.-C. Chu$^{(1)}$, J.~M.~Grandy$^{(2)}$, S.~Huang$^{(3)}$ and
J.~W.~Negele$^{(2)}$}
\vskip 12pt

\baselineskip 10pt
{\narrower\narrower\eightrm
\item{$^{(1)}$}Kellogg Laboratory, California Institute of Technology,
106--38, Pasadena, California\ \ 91125\ \ \ U.S.A.
\medskip
\item{$^{(2)}$}Center for Theoretical Physics, Laboratory for Nuclear Science
and Department of Physics, Massachusetts Institute of Technology, Cambridge,
Massachusetts\ \ 02139\ \  \ U.S.A.
\medskip
\item{$^{(3)}$}Department of Physics, FM--15, University of Washington,
Seattle, Washington\ \ 98195\ \ \ U.S.A.
\medskip}
\vskip 1.5in

\baselineskip 12pt
\centerline{Submitted to: {\it Physical Review Letters}}
\vfill
\centerline{ Typeset in $\TeX$ by Roger L. Gilson}
\vskip -12pt
\noindent CTP\#2113\hfill June 1992
\eject
\baselineskip 24pt plus 2pt minus 2pt
\centerline{\bf ABSTRACT}
\medskip
The first exploratory calculations of QCD vacuum correlation functions on a
lattice are reported.  Qualitative agreement with phenomenological results is
obtained in channels for which experimental data are available, and these
correlation functions are shown to be useful in exploring approximations based
on sum rules and interacting instantons.

\vskip .4in
\centerline{PACS numbers: 12.38.Gc}
\vfill
\eject
Vacuum correlation functions for space-like separated hadron currents provide a
useful tool for exploring the structure of the QCD vacuum which supplements our
knowledge of hadronic ground states.  These correlation functions have been
determined phenomenologically in some channels
by using dispersion relations to analyze hadron
production and $\tau$-decay
experiments,$^1$ and are also amenable to calculations using lattice
field theory, QCD sum rules, and the interacting instanton
approximation.  Hence, in this work, we report the first exploratory lattice
QCD calculation of meson and baryon correlation functions to demonstrate
reasonable agreement with experimentally measured results where available
and to show the
potential more extensive calculations have to explore, test, and refine
approximations.

The lattice calculations were performed on a $16^3\times 24$ lattice in the
quenched approximation at inverse coupling
$6/g^2 = 5.7$, corresponding to a physical lattice
spacing defined by the proton mass of approximately $0.168~$fm.
This inverse coupling constant is large enough to give a semi-quantitative
approximation and has the significant advantage that light quark propagators
are available for point sources.  Hence,
propagators with point sources calculated
by Soni {\it et al.\/}$^{2}$ for 16 configurations with five values of
the quark mass $1/2\kappa - 1/2\kappa_c$ ranging from 351~MeV to 25~MeV were
used to calculate the meson and baryon correlation functions listed in
Table~I.  In order to calculate the ratio of interacting to free correlation
functions, corresponding
free lattice propagators at a very small quark mass, $ma = 0.05$ were
calculated exactly on a $(48)^4$ lattice, where the lattice volume was
chosen large enough to eliminate finite volume effects for spatial separation
less than 2~fm.  Because the propagators had hard-wall boundary conditions in
the time direction, all correlation functions were calculated on the central
time slice containing the source.

Several lattice artifacts had to be eliminated to obtain a physical
approximation to the continuum correlation functions.  Significant anisotropy
is introduced in the rotationally invariant continuum correlation functions by
the Cartesian lattice.  Although at very short physical distances the
anisotropies in the numerator and denominator cancel by asymptotic freedom and
at a sufficiently large number of lattice spacings the granularity of the
lattice becomes negligible, for the present lattice spacing there remains
an intermediate region in which some correlation function ratios display
substantial anisotropy.  As discussed and justified in detail in a subsequent
article,$^3$ since
the lattice points far away from the Cartesian directions
 agree with the continuum result in the non-interacting
case and should be the most reliable in the interacting theory, the lattice
results in this work are taken from sites ${\svec r}$
within a small angular cone surrounding
the diagonal directions ${\svec d} = (n,n,n)$
such that $\hat d \cdot
\hat r \ge 0.9$.  Also, although in principle one would like to normalize
the ratios at an infinitesimally small separation, we have had to normalize
our results at the first non-zero diagonal separation $(1,1,1)$, corresponding
to a physical separation of $\sim 0.29$~fm.

A second lattice artifact arises from the contributions of periodically
repeated images of the physical sources on the finite lattice having periodic
boundary conditions.  As discussed in detail in Ref.~[4], it is
straightforward to perform a self-consistent subtraction of the image
contributions, and we have applied and verified this image correction to all
the correlation functions presented in this work.

Finally, because it is impractical to calculate quark propagators at a quark
mass light enough to produce a physical pion, it is necessary to perform a
sequence of calculations at a series of heavier quark masses and extrapolate
to the pion limit.  Although we do not know the proper functional form for the
extrapolation in the chiral limit, the extrapolation is in fact innocuous
because we only have to extrapolate data at $m_q = 351$, 199, 110, 67, and
25~MeV down to 8~MeV.  A worst case example will be shown below in Fig.~1.
Thus, we believe the technical aspects of relating the present lattice
calculations to the physical continuum theory are under reasonable
control, and that our
results provide a meaningful first
comparison of quenched QCD with phenomenological and model results.

  The correlators we considered are listed in Table~I. For
convenience, we have contracted the vector indices in the vector
and axial channels. In addition, we inserted a factor of
$x_\mu \gamma^\mu$ in the baryonic correlators before taking
the Dirac trace in order to project out the component which
is stable in the massless quark limit.
For a local field theory such as QCD, a two-point
function is uniquely characterized by its absorptive part in
momentum space (up to a polynomial) through the dispersion
relation
$$\int d^4x\, e^{iqx} \ll0\left|T J(x) J(0) \right|0\rr
  = \left\{\matrix{ 1\cr 3q^2\cr-iq^\mu\gamma_\mu\cr}\right\}
\left( \int ds {f(s)\over s-q^2} + P(q^2)\right) +\ldots\eqno(1)$$
where the upper, middle and lower terms in brackets refer to scalar and
pseudoscalar mesons, vector and axial vector mesons and baryon
channels respectively and terms which vanish for the correlation
functions in Table~I have been omitted.
Phenomenologically, we expect $f(s)$ has two major contributions,
a resonance piece and a continuum piece, parameterized as
$f(s)=\lambda^2\delta(s-M^2)+f_p(s)\theta(s-s_0)$,
where $M$ is the resonance mass, $\lambda$ is the coupling of the
current to the resonance state and $f_p(s)$ is the perturbative
contribution to the correlator. Since QCD is asymptotically free,
the $f_p(s)$'s are well-approximated by the corresponding free
correlators with massless quarks, also listed in Table I. This
type of parameterization is widely used in QCD sum rule
calculations.

  An inverse Fourier transform of Eq.~(1) defines the
following phenomenological correlators
in coordinate space as a function of $M$, $\lambda$ and $s_0$,
where the polynomial $P(q^2)$ only contributes at the point
$x=0$ and can be ignored for finite $x$:
$$R^{\left\{ {{m}\atop {b}}\right\}}
(x) = \lambda^2M \left\{ \matrix{
x^{-1}\cr 3M^2x^{-1}\cr M\cr}\right\} K_{\left\{ {{\scriptstyle
1\atop\scriptstyle 1}\atop \scriptstyle 2}\right\} }
(Mx) + \int^\infty_{s_0} ds\, f_p (s) \left\{ \matrix{ \sqrt{s}\,x^{-1}\cr
3s^{3/2} x^{-1}\cr
s\cr}\right\} K_{\left\{ {{\scriptstyle 1\atop \scriptstyle 1}\atop
\scriptstyle
2}\right\}}
\left( \sqrt{s}\,x\right) \eqno(2)$$
Here, $m$ refers to the upper terms in brackets for scalar and pseudoscalar
mesons and the middle terms in brackets for vector and axial vector mesons,
and $b$ refers to the lower terms in brackets for baryon channels.
In practice, we always normalize $R(x)$ by the corresponding
free correlator with massless quarks $R^m_0(x)\propto x^{-6}$
and $R^b_0(x)\propto x^{-8}$.

  In Fig.~1 we show the complete lattice data and the
three-parameter fit in the pseudoscalar meson (pion) channel for each of
five values of the bare quark mass. The result
at the physical pion mass is obtained by binning the lattice data in
two-lattice-unit bins and extrapolating the binned data with the results shown
by the solid circles.  The striking result in this channel is the extremely
rapid rise in the correlation function ratio, necessitating a log plot, which
arises from the strong attraction and corresponding light pion mass.  The
lattice result agrees qualitatively with Shuryak's phenomenological
estimate,$^1$ denoted by the dot-dashed line, based on the value $\lambda_\pi
= (480~\hbox{MeV})^2$ and the fact that
the peak is proportional to $\lambda_\pi/m^2_\pi \sim f_\pi/m_q$.
explains the particularly large quark mass dependence in this
channel.  Detailed treatment of the extrapolation, error analysis of fitted
parameters, and lattice renormalization corrections will be deferred to the
longer paper,$^3$ and we only show extrapolated results and phenomenological
fits
for all other channels, where the quark mass dependence is much weaker.

Figure 2a displays the vector meson channel (rho) result.  As emphasized by
Shuryak, the salient feature in this channel is the fact that although the
free correlator falls four orders of magnitude between 0.3 and 1.5~fm, the
ratio is nearly one over the whole range, and our lattice result is consistent
with his phenomenological analysis of $e^+ e^-\to$ even number $\pi$'s,
denoted by the dashed curve.  The result in the axial meson channel, ($A_1$)
shown in Fig.~2b is qualitatively similar to the phenomenological analysis of
$\tau\to 3\pi$ decay$^1$ denoted by the dashed line, although finite lattice
effects render it difficult to reproduce the rising tail due to mixing with
the pion.  The result for the scalar meson channel, for which the
extrapolation was more problematic than any other, is shown in Fig.~2c.  Note
that in
both the axial and scalar channels, Eq.~(2) did not produce reasonable
physical parameters when fit to the data, so
the solid curves
are smooth curves to guide the eye in these cases. For
comparison, the predictions of the interacting instanton approximation$^5$ for
mesons using a Pauli--Villars cutoff $\Lambda_{\rm PV} = 130$~MeV are shown by
dotted lines, and in each case are in qualitative agreement with the lattice
results.  Comparison of calculations on the same lattice with
``cooled'' configurations which are in progress
will be particularly instructive in this connection.

Since there are no phenomenological data in the baryon channels,
in Fig. 3 we have
compared nucleon and delta correlators with the sum rule calculations of
Belyaev and Ioffe$^6$ (dashed lines) and Farrar {\it et al.\/}$^7$ (dot-dashed
lines).  Given the agreement with phenomenology in the meson channels and
drastic differences in sum rule results for the delta, lattice correlation
functions can play a useful role in testing and refining sum rule
approximations.  In summary, we believe these exploratory calculations
demonstrate the feasibility and utility of calculating vacuum correlation
functions on a lattice and motivate definitive calculation on larger
lattices with $6/g^2\ge 6$.

It is a pleasure to thank Edward Shuryak for stimulating discussions at the
Aspen Center for Physics and elsewhere and for making his data available to us
prior to publication.  We also thank
Amarjit Soni for making his point propagators
available to us, and the National Energy Research Supercomputer Center for
Providing Cray-2 computer resources.
\vfill
\eject
\centerline{\bf REFERENCES}
\medskip
\item{1.}E.~Shuryak, Stony Brook preprint SUNY-NTG-91/45,
to appear in {\it Rev. Mod. Phys.\/} (1992).
\medskip
\item{2.}A.~Soni, {\it National Energy Research Supercomputer Center
Buffer} {\bf 14}, 23 (1990).
\medskip
\item{3.}M.-C. Chu, J.~M.~Grandy, S.~Huang and J.~W.~Negele, to be published.
\medskip
\item{4.}M.~Burkardt, J.~M.~Grandy and J.~W.~Negele, MIT preprint CTP\#2108
(1992).
\medskip
\item{5.}E.~Shuryak, {\it Nucl. Phys.\/} {\bf B328}, 102 (1989).
\medskip
\item{6.}B.~L.~Ioffe, {\it Nucl. Phys.\/} {\bf B188}, 317 (1981); V. M.
Belyaev and B. L. Ioffe, {\it Sov. Phys. JETP} {\bf 83}, 976 (1982).
\medskip
\item{7.}G.~Farrar, H.~Zhoang, A.~A.~Ogloblin and I.~R.~Zhitnitsky, {\it Nucl.
Phys.\/} {\bf B311}, 585 (1981).
\vfill
\eject
\centerline{\bf FIGURE CAPTIONS}
\medskip
\item{Fig.~1:}Ratio of interacting to free correlation functions measured at
five quark masses (open circles), binned data extrapolated to the physical
pion mass (closed circles), three-parameter fits (solid lines) and the
phenomenological results of Ref.~[1] (dot-dashed line).
\bigskip
\item{Fig.~2:}Extrapolated ratio of meson correlation functions (closed
circles) and
fits (solid curves) as in Fig.~1.
Dashed lines denote phenomenological results$^1$ and dotted
lines show the interacting instanton approximation.$^5$
\bigskip
\item{Fig.~3:}Extrapolated ratio of baryon correlation functions (closed
circles) and fits (solid curves) as in Fig.~1.  Dashed and dot-dashed lines
denote sum rule results from Ref.~[6] and [7], respectively.
\vfill
\eject
$$\hbox{\vbox{\offinterlineskip
\def\superstrut{\hbox{\vrule height 20pt depth 15pt width 0pt}}
\def\strut{\hbox{\vrule height 12pt depth 6pt width 0pt}}
\hrule
\halign{
\strut\vrule#\tabskip 0.05in&
#\hfil &
\vrule# &
$#$\hfil &
\vrule# &
\hfil$#$\hfil &
\vrule# &
\hfil${\displaystyle{#}}$\hfil &
\vrule#\tabskip 0.0in\cr
&\multispan7\hfil{\bf Table I: Hadron currents and correlation functions.}
\hfil & \cr\noalign{\hrule}
& \omit\hfil Channel\hfil && \omit\hfil Current\hfil && \omit\hfil
Correlator\hfil && f_p(s) & \cr\noalign{\hrule}
\superstrut& Vector && J_\mu = \bar{u} \gamma_\mu d && \ll 0 \left| T J_\mu(x)
\bar{J}_\mu(0) \right|0\rr && {1\over 4\pi^2} & \cr\noalign{\hrule}
\superstrut
& Axial && J^5_\mu = \bar{u} \gamma_\mu \gamma_5 d && \ll 0\left|TJ^5_\mu(x)
\bar{J}^5_\mu(0) \right|0\rr && {1\over 4\pi^2} & \cr\noalign{\hrule}
\superstrut
& Pseudoscalar && J^p = \bar{u} \gamma_5 d && \ll 0 \left| T J^p(x)
\bar{J}^p(0) \right|0\rr && {3s\over 8\pi^2} & \cr\noalign{\hrule}
\superstrut
& Scalar && J^s = \bar{u}d && \ll 0 \left| T J^s (x) \bar{J}^s(0)
\right| 0 \rr &&
{3s\over 8\pi^2} & \cr\noalign{\hrule}
\superstrut& Nucleon && J^N = \epsilon_{abc} \left[ u^a C \gamma_\mu u^b\right]
\gamma_\mu\gamma_5 d^c && {\displaystyle{1\over 4}} \Tr \left( \ll 0 \left| T
J^N (x) \bar{J}^N (0) \right| 0 \rr x_\nu \gamma^\nu\right) && {s^2
\over 64\pi^4} & \cr\noalign{\hrule}
\superstrut
& Delta && J^\Delta_\mu = \epsilon_{abc} \left[ u^a C \gamma_\mu u^b \right]
u^c && {\displaystyle{1\over 4}} \Tr \left( \ll 0 \left| T J^\Delta_\mu(x)
\bar{J}^\Delta_\mu(0) \right| 0 \rr x_\nu \gamma^\nu\right) && {3s^2
\over 256 \pi^4} & \cr\noalign{\hrule}}}}$$
\par
\vfill
\end